\documentstyle[prl,aps,epsf]{revtex}
\draft
\begin{document}
\twocolumn[\hsize\textwidth\columnwidth\hsize\csname @twocolumnfalse\endcsname

\title{Ratchet Effect and Nonlinear Transport for Particles on Random Substrates with Crossed ac Drives}     
\author{C. Reichhardt and C.J. Olson Reichhardt} 
\address{ 
Center for Nonlinear Studies and Theoretical 
Division, 
Los Alamos National Laboratory, Los Alamos, New Mexico 87545}

\date{\today}
\maketitle

\begin{abstract}
We show in simulations that overdamped interacting 
particles in two dimensions with a 
randomly disordered substrate can exhibit 
novel nonequilibrium transport phenomena including a 
transverse ratchet effect, where 
a combined dc drive and circular ac drive produce
a drift velocity 
in the direction transverse to the applied dc drive.  
The random disorder does not 
break any global symmetry; however, in two dimensions,
symmetry breaking occurs due to the chirality of the circular drive. 
In addition to inducing the transverse 
ratchet effect, increasing the ac amplitude also strongly
affects the longitudinal velocity response and can produce
what we term an overshoot effect where the
longitudinal dc velocity is higher in the presence of the ac drive than 
it would be for a dc drive alone.   
We also find a dynamical reordering transition upon
increasing the ac amplitude.
In the absence of a dc drive, it is possible to 
obtain a ratchet effect when 
the combined ac drives produce particle orbits that break a 
reflection symmetry. 
In this case, as the 
ac amplitude increases, current reversals can occur.
These effects may be observable 
for vortices in type II superconductors as well as for colloids interacting
with random substrates.  
\end{abstract}

\pacs{PACS numbers: 05.60.-k, 05.45.-a, 74.25.Qt}
\vskip2pc]
\narrowtext

\section{Introduction}
There has been growing interest 
in studying nonequilibrium transport of particles
on asymmetric substrates in which a dc flow of particles
can arise from the application of an ac drive or by flashing
the substrate on and off periodically \cite{Review}.
This rectification
phenomena is often called a ratchet effect and has been studied
in the context of molecular motors \cite{Rachet}, colloidal 
matter \cite{Grier}, transport of atoms in 
optical traps \cite{Atom}, granular matter \cite{Det},
electron transport in asymmetric geometries \cite{Linke}, 
vortex transport and manipulation in type-II superconductors 
\cite{FluxRatchet,Hastings,Olson,Marchesoni,Wordenweber,Moshchalkov},
and transport in Josephson junctions \cite{Falo,Majer}. 
In most of these systems   
the dc transport arises due to
the symmetry breaking caused by an 
imposed one-dimensional asymmetric substrate, such as an asymmetric sawtooth
potential. 
In higher dimensions, there are a larger number of possible ways in which
symmetry can be broken, so it is possible to induce ratchet effects
even when the
substrate is symmetric. 

Recently it was shown that a driven overdamped
particle moving in a two-dimensional periodic substrate can exhibit 
a transverse ratchet effect when driven with both a dc and a circular
ac drive \cite{Reichhardt}.
In this case, the chirality of the ac drive breaks a symmetry
and the dc drive breaks a reflection symmetry, so that the combined
effect is to induce an additional dc particle
current in the direction 
{\it perpendicular} to the applied external dc drive. 
Due to the periodicity of the substrate and the periodicity of the ac driving,
a series of phase locking steps appear in both the
longitudinal velocity and the transverse velocity. 
Even more complicated transport phenomena occur when the ac drive
is not circular but elliptical \cite{Transverse}. 
In this case it is possible to
observe a phenomenon called absolute transverse mobility, where the
particle moves only perpendicular to an applied dc drive. 

Recently, a rich variety of ratchet dynamics were shown to occur for
atoms driven over a two-dimensional substrate with crossed or biharmonic 
ac drives \cite{Guantes}.   
It has also been shown that a ratchet effect can be produced
even in the absence of a dc drive 
for an overdamped particle in 
a two-dimensional periodic potential when the crossed ac drives
produce a particle orbit that breaks a spatial 
reflection symmetry \cite{Transverse}.   
In this case a series of phase locked regions appear as a function
of ac amplitude. In all these cases the substrate is periodic
and the 
symmetry breaking required to produce the ratchet effect arises from the
ac drives. An open question is whether a net dc current 
or ratchet effect can occur for driven particles
on {\it random} substrates when subjected to a circular ac drive. 
Large scale vorticity patterns were 
observed in simulations of particles interacting
with random substrates 
when an ac drive was applied in only one direction \cite{Derenyi}.
The large scale flow arises since random substrates can cause a local 
symmetry breaking, giving rise to a local ratchet effect.  There is no
net global ratchet effect since the 
symmetry is restored on average at large length scales. 
These results suggest that ratchet effects should be
possible in dimensions higher 
than one even when the substrate is random. 

A ratchet effect that occurs on random substrates 
could be of great practical
importance since many systems in which applications for a ratchet effect have
been proposed contain intrinsic random disorder.  If a ratchet effect
could be produced directly from the random disorder,
additional fabrication of structured substrates would 
not be needed. 
One example of such a system is vortices in type II superconductors, 
where random defects act as pinning sites. 
Ratchets previously proposed for this system 
have all included periodic substrates of
some form \cite{FluxRatchet,Hastings,Olson,Marchesoni,Wordenweber,Moshchalkov};
however, the random substrate ratchet effect we study here 
could be achievable experimentally without nanostructuring
the superconducting surface.  

In this work we show that several types of ratchet and nonequilibrium transport
phenomena can occur for particles interacting with a two-dimensional
random substrates with crossed ac driving forces. In the 
absence of the disordered
substrate there is no ratchet effect. For the case where there is a circular
ac drive and an applied dc drive, a transverse ratchet effect can occur 
in which
a net dc drift arises in the direction perpendicular to the
dc drive. If the ac drives are chosen such that
the particles undergo spatially asymmetric 
orbits, directed transport can occur even the absence of the applied dc drive. 
It is also possible to have flux reversals as a function of ac amplitude. 
In the absence of disorder, the 
multiple ac drives do not produce a ratchet effect. 
We specifically demonstrate these ratchet effects for 
vortices in type-II superconductors
and colloidal particles confined to two dimensions.  

\section{Simulation}

We consider an assembly of $N_v$
overdamped interacting particles in two dimensions, and 
impose periodic boundary conditions
in the $x$ and $y$ directions. 
The equation of motion for an individual particle
$i$ is
\begin{equation}
\eta{\bf v}_i = {\bf f}_{i} =  
{\bf f}^{i}_{int}+{\bf f}^{i}_{p}+{\bf f}_{dc}+{\bf f}_{ac} +{\bf f}_{i}^{T} 
\end{equation}
where  the damping constant
$\eta$ is set to unity. 
The equation of motion is integrated according to the leap-frog method with a 
normalized time step $dt = 0.001$. 
The interaction force from the other particles is 
${\bf f}^{i}_{int} = -\sum^{N_v}_{j \neq i}\nabla_i  U(r_{ij})$,
where $r_{ij}$ is the distance between particles $i$ and $j$. 
We consider two forms of particle-particle interactions. 
Vortices in a thin film type II superconductor have a repulsive
interaction with $U(r_{ij}) = -\ln(r_{ij})$ which gives a long
range repulsive force $1/r_{ij}$. We have previously simulated vortices
in periodic arrays of pinning sites using this interaction
and a similar equation of motion. In order to treat
the long-range interactions efficiently, 
we employ a fast summation technique \cite{Niels}.   
Additionally we have considered a short range
repulsive Yukawa interaction
with inverse screening length $\kappa$,  
$U(r_{ij}) = \exp(-\kappa r_{ij})/r_{ij}$, 
which is appropriate for colloidal particles. 
In this case, we assume that we are working in the strongly charged,
low volume fraction limit, so that hydrodynamic interactions can
be neglected.

The  force from the random substrate is modeled
as $N_{p}$ randomly placed attractive parabolic wells with

\begin{figure}
\center{
\epsfxsize=3.5in
\epsfbox{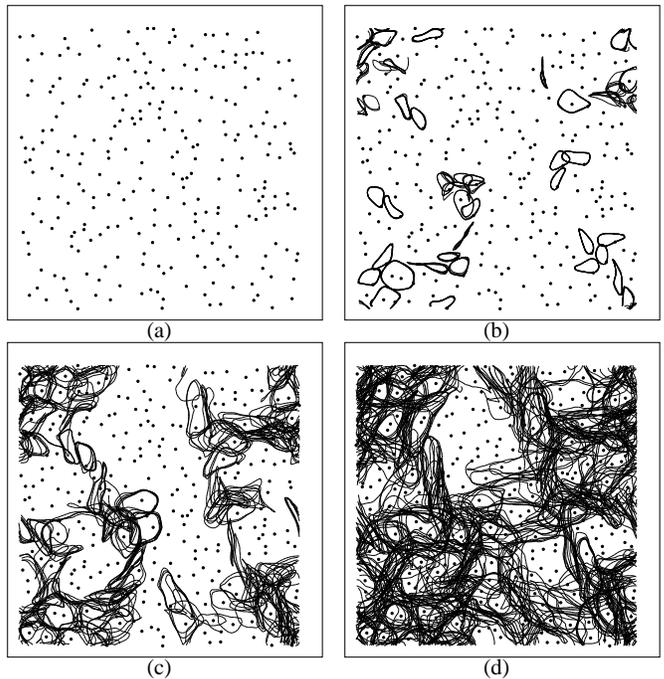}}
\caption{
Particles (black dots) and trajectories (black lines) for a fixed 
period of time for a system with random pinning and a 
circular ac drive with fixed frequency $\omega=0.001$ and increasing
amplitude $A$. (a) $ A = 0.0$, (b) $A = 0.5$, (c) $A = 0.8$, and
(d) $A = 1.2$.     
}
\end{figure}

\begin{equation}
{\bf f}_{p}^{i} = -\sum_{k=1}^{N_p}
(f_{p}/r_{p}) |{\bf r}_{i} - {\bf r}_{k}^{(p)}|\
\Theta(r_{p} - |{\bf r}_{i} - {\bf r}_{k}^{(p)}|)
{\hat {\bf r}_{ik}^{(p)}}   
\end{equation} 
where $\Theta$ is the Heaviside step function, ${\bf r}_{k}^{(p)}$ is 
the location of pinning site $k$, $f_{p}$ is the maximum pinning force,
$r_{p}$ is the pinning site radius and 
${\hat {\bf r}}_{ik}^{(p)} = 
({\bf r}_{i} - {\bf r}_{k}^{(p)})/|{\bf r}_{i} - {\bf r}_{k}^{(p)}|$.   
The force from the temperature ${\bf f}_i^{T}$
is modeled as random Langevin kicks with the properties 
$\langle f^{T}_i(t)\rangle = 0$ 
and $\langle f^{T}_i(t)f^{T}_j(t^{\prime})\rangle = 2\eta k_{B}T
\delta_{ij}\delta(t -t^{\prime})$. 
Except where noted, the results are obtained at $T=0$.
In some cases we apply a dc force, ${\bf f}_{dc}$, which is modeled as a 
uniform force on all the particles in the $x$ direction. 
In this work we increase the 
dc force in small increments, and we have checked that the force increase 
is small enough that transient effects are negligible. 
The ac drive force is 
\begin{equation} 
{\bf f}_{ac} = f(t){\hat {\bf x}} + 
g(t){\hat {\bf y}}.
\end{equation}
Here $f(t)$ and $g(t)$ are oscillating functions with 
$\langle f(t)\rangle_{\tau} = 0$ and 
$\langle g(t)\rangle_{\tau} = 0.0$ over a period $\tau$.  
In the first part of this work we will consider the
circular drive  
case of $ f(t) = A\sin(\omega t)$ and $g(t) = A\cos(\omega t)$. 
We measure the particle trajectories and velocities
in the longitudinal   
$\langle V_{x}\rangle = 
(1/N_v)\langle\sum_{i}^{N_{v}}{\hat {\bf x}}\cdot {\bf v}_{i}\rangle$ and
transverse direction
$\langle V_{y}\rangle = 
(1/N_v)\langle\sum_{i}^{N_{v}}{\hat {\bf y}}\cdot {\bf v}_{i}\rangle$. 
We average over several hundred periods of the ac drive in 
order to obtain a steady state average.   

\begin{figure}
\center{
\epsfxsize=3.5in
\epsfbox{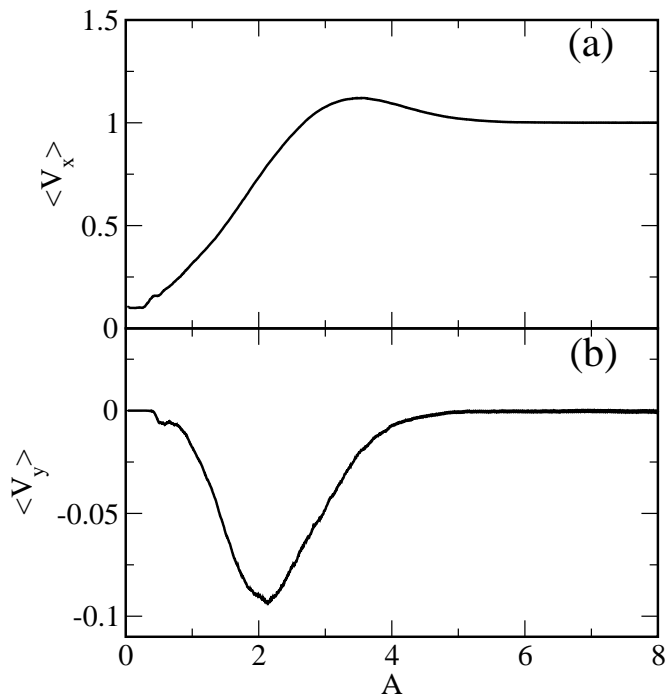}}
\caption{
The average velocity response in the (a) $x$ and (b) $y$ directions for
a system with 
$f_p=3.5$, a fixed dc drive of $f_{dc} = 1.0$, and increasing
circular ac drive amplitude $A$. 
}
\end{figure}

\section{Circular ac Drives and Transverse Rectification}

\subsection{Transverse and Longitudinal Velocity-Force Characteristics}    

We first consider the case of a circular ac drive
and no dc drive. 
In Fig.~1 we illustrate the particle motion 
over a fixed period of time for a system with 
$N_{p} = 250$, $N_{v} = 280$, $f_p=3.5$, $T = 0.0$,
$f_{dc}=0$,
fixed frequency $\omega=0.001$, and an increasing ac amplitude $A$. 
In Fig.~1(a), $A = 0.0$ and all the particles are stationary. In Fig.~1(b), 
at $A = 0.5$, a portion of the particles are pinned and do
not move with the ac drive; however, several particles are now 
mobile and follow closed paths. In the absence of other particles the
paths would be circular, but due to the repulsion of the neighboring
trapped particles, the paths are strongly distorted from circular
shapes.
In general, at low ac amplitude some of the particles move in 
closed paths but 
there is no long time particle diffusion. 
If there are many more particles than
pinning sites, $N_v\gg N_p$, 
it is possible for meandering paths that
change with time to form, permitting some particles to diffuse throughout the
entire sample over long time scales. When the number of particles is close
to or less than the number of pinning sites, 
$N_v\lesssim N_p$, 
similar meandering paths can only form at 

\begin{figure}
\center{
\epsfxsize=3.5in
\epsfbox{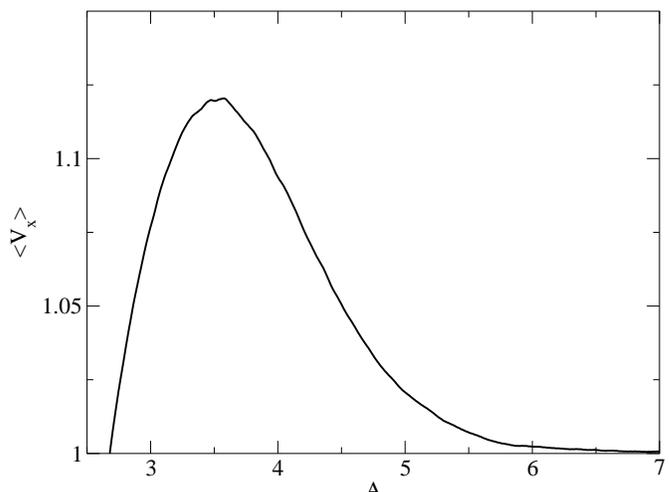}}
\caption{
A closeup of $\langle V_{x}\rangle$ vs $A$ 
for the system in Fig.~2(a) showing the
overshoot region where 
$\langle V_{x}\rangle>1.0$.
}
\end{figure}

\noindent
higher ac amplitudes
when pinning and repinning of particles at the pinning sites 
becomes possible, as shown
in Fig.~1(c) for $A = 0.8$. As the
ac amplitude is further increased, more of the particles become mobile
and eventually all the particles depin. In Fig.~1(d) we show
the case for $A = 1.2$ where large portions of the particles are depinned
and the channel structures 
in which motion is occurring change rapidly with time. 

In the absence of a dc drive, 
a purely circular ac drive does not 
produce a net drift velocity, and 
$\langle V_{x}\rangle = \langle V_{y}\rangle = 0$.   
If the system size is very small and finite
size effects become relevant, it is possible to 
observe a small dc drift.
For the system sizes we consider here, these 
finite size effects are eliminated.
In Fig.~2 we plot $\langle V_{x}\rangle $ and $\langle V_{y}\rangle$
vs $A$ for the 
same system in Fig.~1 with a fixed dc drive $f_{dc}=1.0$ applied in the $x$ 
direction. In the absence of pinning, 
a single particle subjected to this combination of drives
would move in the $x$-direction
with $\langle V_{x}\rangle = 1.0$. 
For $A = 0$, most of the particles are pinned,
$\langle V_{x}\rangle = 0.1$, and 
$\langle V_{y}\rangle = 0$. For $A < 0.58$,  there is little
change in $\langle V_{x}\rangle$ and 
$\langle V_{y}\rangle$, while for $0.58 < A < 2.0$
the velocity in the $x$-direction  
increases, indicating that more particles are depinning and
moving in the direction of the dc drive. Over this same range
of $A$,
$\langle V_{y}\rangle$ drops below zero and develops an increasingly
negative value,
indicating that particles are drifting
in the negative $y$-direction
in spite of the fact that there
is no applied dc drive in the $y$ direction. 
The transverse velocity reaches a 
maximum magnitude $|\langle V_{y}\rangle| = 0.09$ 
near $A=2.1$ and then gradually returns to zero at $A\sim 5.$
For $A > 6$ the longitudinal velocity saturates at
$\langle V_{x}\rangle = 1.0$  indicating that the pinning has been
washed out and the particles are moving at the same velocity as
they would in the absence of pinning.
For $2.6<A<6$, in a range falling above the maximum of
$|\langle V_{y}\rangle|$ but below the saturation
of $\langle V_{x}\rangle$,
there is what we term an overshoot effect where the
longitudinal velocity 

\begin{figure}
\center{
\epsfxsize=3.5in
\epsfbox{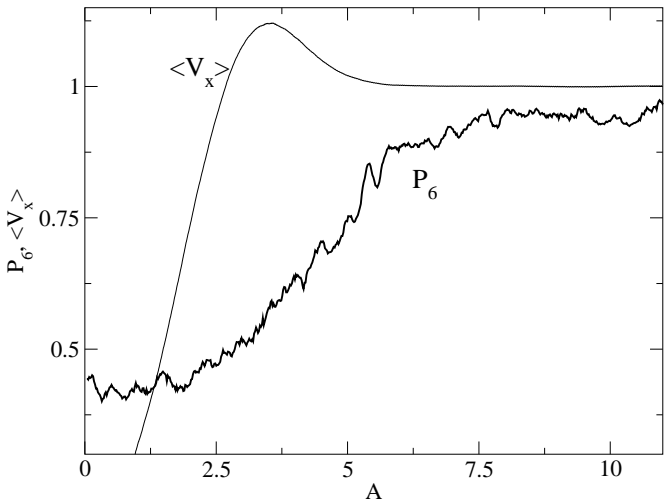}}
\caption{ 
The fraction of six-fold coordinated particles $P_{6}$ vs $A$ for the
same system in Fig.~2.  
$\langle V_{x}\rangle$ vs $A$ is also plotted for direct comparison
to show that the overshoot effect is lost when the
lattice reorders.
}
\label{fig:Matt}
\end{figure}

\noindent
$\langle V_{x}\rangle$ is {\it greater} than the
maximum possible velocity expected for the applied dc drive,
$\langle V_{x}\rangle > 1.0$.
In Fig.~3(a) 
we show a blowup of this region where there is a peak value
of $\langle V_{x}\rangle =1.12$ at $A=3.5$, giving an excess velocity
of $0.12$.
This implies that some of the energy from the $x$ component of the ac drive
is being coupled into the particle motion during the positive $x$ direction 
half of the ac cycle, but an equivalent amount of energy is not
being removed during the negative $x$ direction half of the
ac cycle, producing a net imbalance.
The random substrate plays an important role in this effect, as indicated
by the fact that $\langle V_{x}\rangle$ gradually drops back to 
a saturation value of $1.0$ as the ac drive amplitude is further increased
and washes out the effect of the pinning.

The transverse ratchet and the overshoot 
effect can both be understood as arising from
the combination of the symmetry breaking from the circular drive and the
nonlinear features of the velocity-force curves
that appear when plasticity is induced as particles are driven over
a random substrate, as we will demonstrate in the following sections.
In general, for particles such as vortices or colloids  
moving over strong random disorder
under the influence of a dc drive $f$, there
are three distinct dynamic regimes \cite{Higgins,Vinokur,Zimanyi,Balents}.
For low drives $f$, 
all the particles are pinned, and there is a
threshold depinning force $f_{c}$ 
that must be applied before motion begins. 
As the driving is 
increased further above depinning,  
$f\gtrsim f_c$, there is a plastic flow regime where 
pinned and flowing particles coexist.
In this regime 
the velocity vs force curves are non-linear and 
can be fit to a power law form
$V = (f - f_{c})^{\beta}$.  
For elastic depinning, $\beta=2/3$, but for the plastic depinning
that occurs for strongly disordered substrates,
$\beta >1.0$ \cite{Olson,Fisher,Marchetti}. 
At even higher drives, a dynamical
reordering effect occurs in which the particles 

\begin{figure}
\center{
\epsfxsize=3.5in
\epsfbox{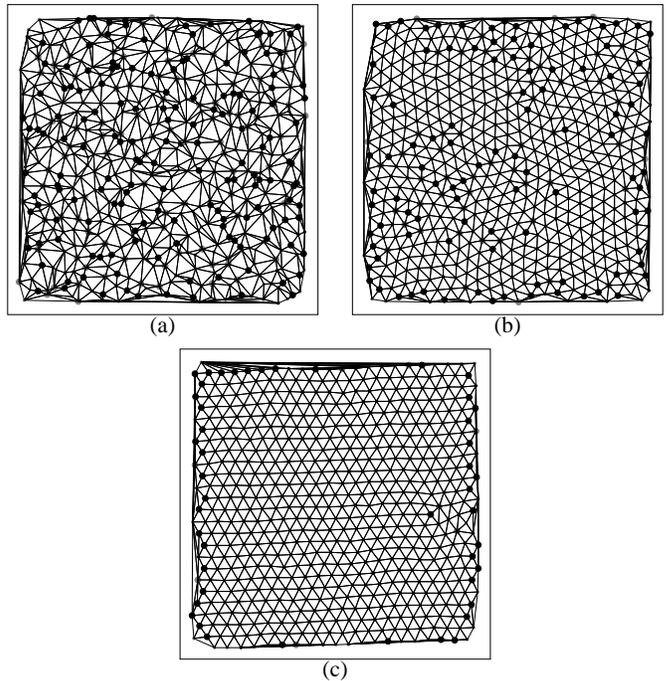}}
\caption{
The Delaunay triangulation of the particle positions 
at one instant
for different values of $A$ for the system in Fig.~4. 
The large black dots indicate fivefold coordinated particles
and the lighter large dots indicate sevenfold coordinated particles.
(a) $A = 0.5$, (b) $A = 2.5$, and (c) $A = 10$.    
}
\end{figure}

\noindent
crystallize or
partially crystallize when 
the longitudinal component of the pinning is effectively washed out.
Above this 
transition, the velocity force curves are linear.
The reordered state can be a moving crystal or moving smectic state 
\cite{Balents}. 
When these three phases occur, the velocity force curves 
have a distinct S-shape, and the concavity of the
velocity force curve changes from positive to negative somewhere
in the plastic flow regime as the plastic to moving crystal crossover
is approached.
In the crossover regime 
that occurs above the change in concavity, $\beta$ 
drops from $\beta>1$ to $\beta= 1.0$.   

\subsection{Dynamic Reordering with Circular ac Drives}  

If a single ac drive is applied in the $x$ direction instead of a dc drive, 
it is still possible to produce the three 
dynamic regimes described above as a function of ac amplitude and frequency
since the transitions between the regimes are not hysteretic, and 
the system will reorder into a moving smectic state \cite{Bekeris}.
In our system we apply a pair of ac drives to produce a circular ac drive,
rather than a single unidirectional ac drive, and we also add
a dc drive.
We find that a dynamical reordering
transition can still occur in this system, but that
the final reordered state is not 
a moving smectic for small $f_{dc}$
since the ac drive does not select a particular direction.
In Fig.~4 we plot the fraction of six-fold coordinated particles $P_{6}$ vs
$A$ for the same system shown in Figs. 2 and 3, and we also 
superimpose $\langle V_{x}\rangle$ from Fig.~2(a) for a 
direct comparison. 
The quantity $P_{6}$ is obtained  
from a Delaunay triangulation of the system. If the entire lattice is
triangular, then $P_{6} = 1.0$. 
For low $A$ the system is strongly disordered, as indicated by the
fact that $P_{6} \sim 0.4.$ 
As $A$ is increased, $P_{6}$ increases rapidly
toward one for $A > 2.5$, which also corresponds to the region in which
the overshoot effect of $\langle V_{x}\rangle$ occurs. 
For high values of $A>6,$ the 
system is reordered and $P_{6}$ is close to one as the
system enters a moving crystal phase. We note that $P_{6}$
never reaches one due to the boundary effects on 
the algorithm we used to obtain the coordination 
number. 
When $P_{6}$ saturates, the overshoot effect and
the transverse ratchet effect are both lost. 

In Fig.~5 we illustrate the Delaunay triangulations of the particles 
at increasing values of $A$ for the system in Fig. 4.
In Fig.~5(a) for $A = 0.5$, the system is strongly
disordered. 
In Fig.~5(b) for $A = 2.5$, the system is more ordered 
and contains regions of sixfold coordinated particles. 
For $A = 10$, shown in Fig.~5(c), the system is almost
completely reordered into a triangular lattice 
and only a small number of 
dislocations are present. 
We note that the lattice is not 
well aligned with the direction of the dc drive, 
but has its close-packed direction at an angle to the $x$ axis.
The near alignment occurs due to the fact that a small dc drive is
being applied.  When we perform the simulation with zero dc drive,
the reordering still occurs but the direction of the lattice alignment
is random in any given run.
This is in contrast with previous simulations 
performed with dc drives \cite{Zimanyi} where 
the reordered lattices formed moving channels that were 
strongly aligned with the dc drive. 
In the case of reordering induced by a purely dc drive, the dislocations
present at higher drives all have aligned Burgers vectors and the
system forms a smectic state
since the driving force has reduced the pinning only in the direction of
drive, but the pinning transverse to the driving direction still
remains effective \cite{Balents}.
In our ac driven system, the 
one-dimensional channel structures of the smectic state
cannot form since the circular ac motion destroys the effectiveness of
the transverse pinning as well as of the longitudinal pinning.  Thus we
find no moving smectic state.

\subsection{Transverse Ratchet and Overshoot Effect}

Under the circular ac drive, each particle has a nonzero velocity 
component in the positive $y$ direction for half a period, and in the
negative $y$ direction for the other half period.  There is no dc force
applied in the $y$ direction so these components are equal in magnitude.
For the velocity component in the $x$ direction, during half the period
$0<t^* < \tau/2$, where $t^*=t \ {\rm mod}\  \tau$,
the ac and dc drives are in the same direction and the particles
move faster under the maximum force 
$f_{max}=f_{dc}+A$, while in the other half of the
period $\tau/2 < t^* < \tau$,
the two 
drives counteract each other and the particle moves
more slowly under the minimum force $f_{min}=f_{dc}-A$.
If we consider the limit of small ac driving amplitude, where the system 
is still within the nonlinear portion of the velocity-force response curve, 
a portion of the particles have depinned but a portion remain pinned.
Under the drive we apply, the moving particles follow clockwise circles.
During an orbit in which a moving particle encircles a pinned particle, 
the pinned particle exerts a repulsive force on the encircling particle. 
For $0<t^*<\tau/2$, the moving particle passes the pinned particle in a 
short period of time so the repulsive force has less time to deflect the
particle in the $+y$ direction. On the other hand, 
for $\tau/2<t^*<\tau$, the moving particle takes longer to pass 
the pinned particle, so the average deflection in the $-y$ direction is
larger.
As $A$ increases from a small value, more particles become unpinned 
and the average velocity in the $-y$ direction increases. 
When $A$ is large enough that more than 
half the particles are depinned, the $-y$ ratcheting effect 
starts to decrease since there are fewer pinned particles available to 
deflect the paths of the moving particles. 
The velocity should go as the product of the fraction of 
moving particles $n_{m}$ and pinned particles $n_{p}$, 
$\langle V_{y}\rangle =  n_{m}n_{p} = n_{m}(1 - n_{m})$. 
This gives a parabolic shape for $\langle V_y\rangle$,
which is similar to what is observed in Fig.~2(b),           
and the peak should occur when $n_{m} = 1/2$. 
At the larger drives when all the particles are moving and the
velocity-force curve is no longer nonlinear,
the transverse ratcheting effect is lost. The maximum for 
$|\langle V_{y}\rangle|$
in Fig.~2 occurs at $A = 2.0$. If we take the approximation
that the unpinned particles move at the velocity of the applied dc drive,
then $n_m=0.5$ would correspond to $\langle V_{x}\rangle = 0.5$, 
which occurs at $A = 1.5$, not far from the 
maximum in 
$|\langle V_{y}\rangle|$. At high $A$ when 
all the particles are depinned, the transverse drift disappears. 

The overshoot effect can be understood as arising from 
changes in the nonlinear velocity
force curve relations near the reordering transition. Well
below the reordering transition, 
the velocity scales with the driving force as a power law 
$\langle V_{x}\rangle \propto A^{\beta_1}$, 
where $\beta_1>1$, but above the reordering
transition $\langle V_{x}\rangle$ scales linearly with the velocity.
As a result, near but below the reordering transition, the scaling relation
changes form and the scaling exponent may be slightly different, 
giving $\langle V_{x}\rangle \propto A^{\beta_{2}}$,
with $\beta_2 \rightarrow 1$ as the reordering transition is approached. 
For $A$ near but below the reordering transition,
when the ac and dc forces sum to $f_{max}$ in the positive $x$ direction 
during the interval $0<t^*<\tau/2$, 
the velocity follows $A^{\beta_2}$. 
During the other half of the cycle $\tau/2<t^*<\tau$, the particle motion
slows under $f_{min}$ and the velocity follows $A^{\beta_1}$.
The difference during a full period is
$\Delta V_x(A)\approx A^{\beta_{1}} - A^{\beta_{2}}$. 
In the high driving regime 
where the velocity response is linear, $\beta_{1} = \beta_{2} = 1.0$, so
$\Delta V_{x} = 0$. Near the ordering transition, 
$\beta_2$ is very close to 1 since it is very 

\begin{figure}
\center{
\epsfxsize=3.5in
\epsfbox{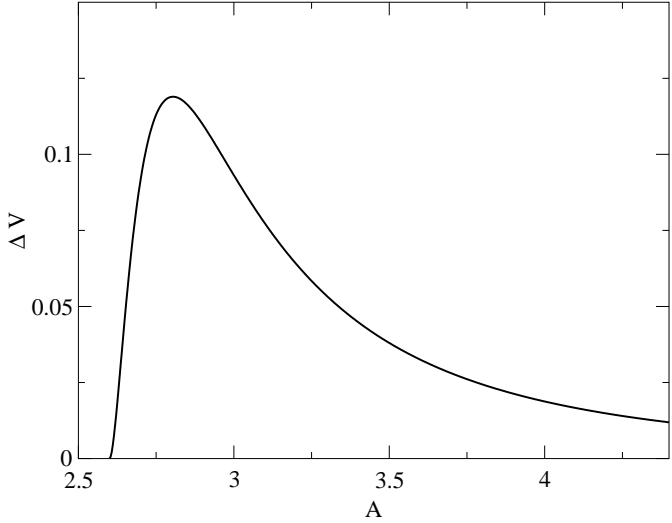}}
\caption{
Plot of the behavior of $\Delta V(A)$ vs $A$ from the phenomenological form 
$\Delta V(A) = B[A^{(1 -(1/A)^{\delta_1})} - A^{(1 - (1/A)^{\delta_2})}]+C$
for $\delta_1 = 4.0$, $\delta_2 = 3.9$, $B = 19.8$,  and $C = 1.6$. 
}
\end{figure}

\noindent
close to the region where
the velocity force curve is linear.
In contrast, $\beta_1>1.0$, so that $A^{\beta_{2}}< A^{\beta_{1}}$.
As a result, there is a net positive contribution from the ac drive to the
dc force over each cycle, 
giving rise to an average velocity $\langle V_{x}\rangle>1$,
higher than the velocity produced by a strictly dc drive. 
We now make a very simple phenomenological assumption for the form
of the excess velocity:
$\Delta V(A) = A^{(1 - (1/A)^{\delta_1})} - A^{(1 - (1/A)^{\delta_2})}$. 
With this form, $\Delta V(A) = 0$ for large $A$ 
when both exponents go to one. 
We also assume that $\delta_{1}$ 
is larger than $\delta_2$ by a small amount, so that 
the exponent on the $\tau/2<t^*<\tau$ half of the cycle is
always smaller than that on the $0<t^*<\tau/2$ half of the cycle.
For all $\delta_{1,2} > 1$ and $A>1$,
$\Delta V(A)$ shows a peak and a slow roll off to zero, 
illustrated in 
Fig.~6, which is 
very similar to the behavior of the excess velocity seen in Fig.~3. 
This suggests
that the overshoot velocity arises 
due to the nonlinear form of the velocity force curves 
near the dynamic reordering transition.

\subsection{Effects of Disorder Strength, 
Particle Density, and Temperature} 

We next examine how the transverse ratchet effect evolves as a function 
of disorder strength. 
In Fig.~7 we plot $\langle V_{y}\rangle$ 
vs $A$ for $f_{p} = 0.5$, 2.5, 3.5, 4.5, and 5.5.  
For the weak disorder $f_p=0.5$, we do not observe any transverse ratcheting,
as indicated by the flat line in Fig.~7. 
As the disorder strength
increases, the maximum value of $|\langle V_{y}\rangle|$ increases 
and the peak value shifts to higher
$A$. The reordering transition at high $A$ also moves out to larger $A$. 
A similar trend appears in the longitudinal velocities 
$\langle V_{x}\rangle$ with the overshoot phenomenon 
showing 

\begin{figure}
\center{
\epsfxsize=3.5in
\epsfbox{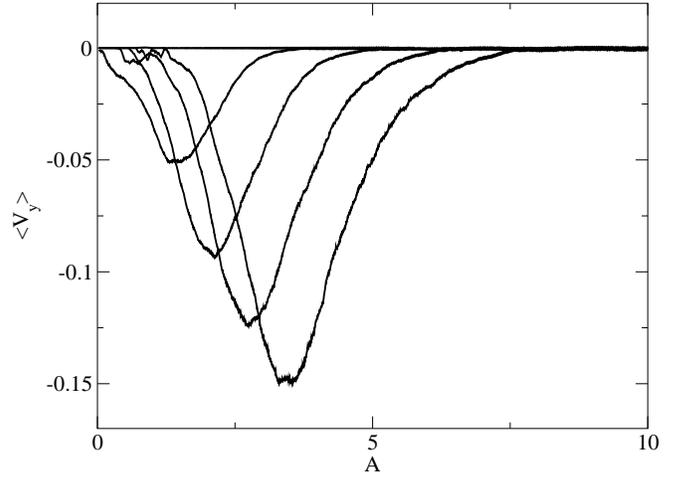}}
\caption{
$V_{y}$ vs $A$ for varied $f_{p}$ for the same system parameters as in Fig.~2.
From top minimum to bottom minimum, $f_{p} = 0.5$
(flat line), 2.5, 3.5, 4.5, and $5.5$.    
}
\end{figure}

\begin{figure}
\center{
\epsfxsize=3.5in
\epsfbox{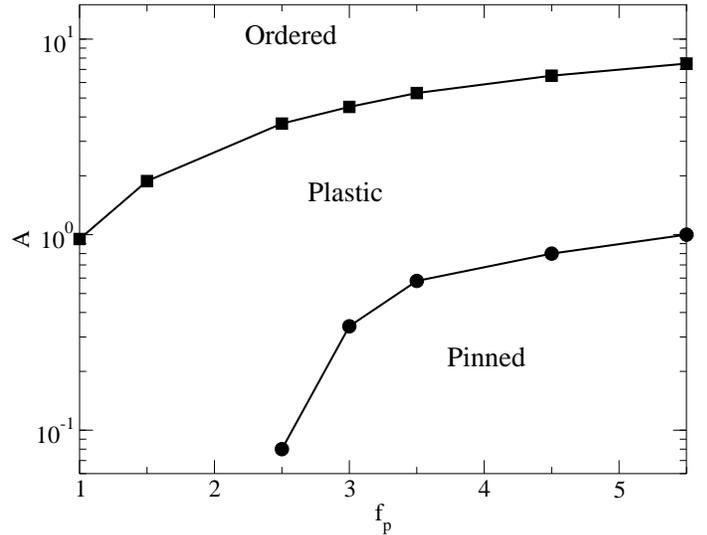}}
\caption{Dynamic phase diagram as a function of $A$ and $f_{p}$
showing the pinned regime, 
plastic flow regime, and reordered regime. The reordered regime is
similar to that illustrated in Fig.~5(c).}
\end{figure}

\noindent
the same behavior as in Fig.~2(a). In Fig.~8 we map out the 
dynamic phase diagram for $A$ vs $f_{p}$. For high $A$ the system forms an
ordered phase similar to that illustrated in Fig.~5(c). For low $A$ and 
high $f_{p}$ the system is in the pinned phase. 
We note that, due to the constant applied $f_d=1$, if 
$f_{p}<f_d$, the system passes directly into a moving phase 
and is never pinned.
The intermediate regime 
is the plastic flow phase where there is a finite transverse velocity.
As $f_{p}$ is lowered, the onset of the ordered phase drops to lower values
of $A$. For $f_{p} < 1.0$ the pinning is weak enough that large portions
of the 

\begin{figure}
\center{
\epsfxsize=3.5in
\epsfbox{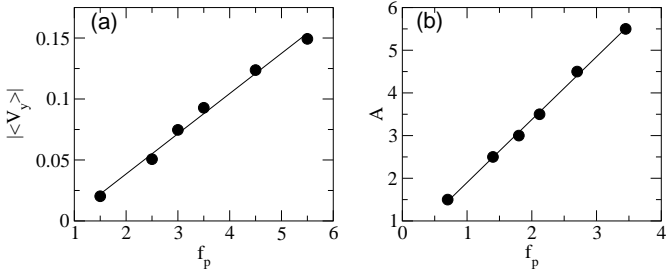}}
\caption{
(a) The maximum value of $|\langle V_{y}\rangle|$ vs 
$f_{p}$ for the system in Fig.~7.
The solid line is a linear fit. 
(b) The value of $A$ at which the maximum 
in $|\langle V_y\rangle|$ occurs as a function of $f_{p}$
for the system in Fig.~7. The solid line is a linear fit.        
}
\end{figure}

\begin{figure}
\center{
\epsfxsize=3.5in
\epsfbox{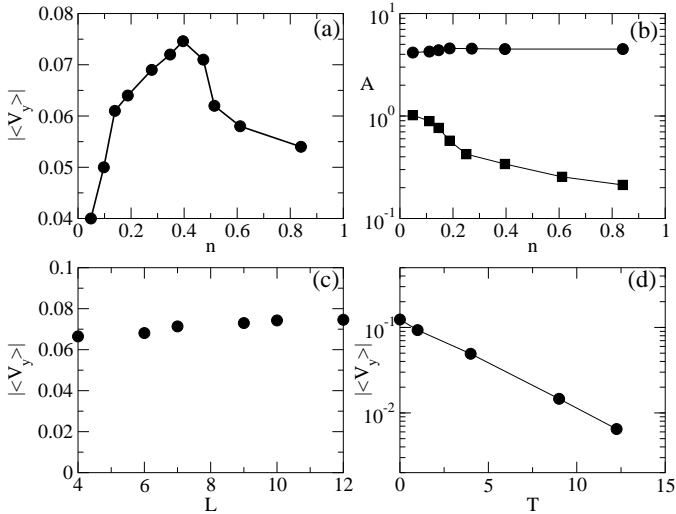}}
\caption{
(a) The maximum value of $|\langle V_{y}\rangle|$ vs 
particle density $n$ for a system with $f_{p} = 3.0$.  
(b) The dynamic phase diagram  for $A$ vs $n$. 
The squares separate the pinned phase from 
the plastic flow phase and 
the solid circles separate the plastic flow phase from the 
reordered moving phase. 
(c) The maximum in $|\langle V_{y}\rangle|$ 
vs system size $L$ for a system with
fixed $n = 0.4$ and $f_{p} = 3.0$. 
(d) The maximum value of of $|\langle V_{y}\rangle|$ vs $T$ for 
a system with fixed $n = 0.3$ and $f_{p} = 3.0$.   
}
\end{figure}

\noindent
lattice start to crystallize and the transverse ratcheting effect
becomes very difficult to detect.   
In Fig.~9(a) we show that the maximum value of 
$|\langle V_{y}\rangle |$ increases
linearly with 
the maximum pinning force and in Fig.~9(b) we show that
the value of $A$ where the peak value in 
$|\langle V_{y}\rangle|$ occurs also increases linearly with $f_{p}$.   

Next we consider the effects of changing the 
particle density $n$. We fix both
the ratio of particles to pinning sites and the system size.
In this way, the effects of the particle-particle interactions can
be studied. The transverse ratchet arises due to the interaction between
the moving particles and the repulsive force from the pinned particles.
If the density of the system is lowered, the average velocity 
$\langle V_y\rangle$ of the transverse ratchet effect
should drop since the particles are further apart. 
In Fig.~10(a) we plot the peak value for 
$|\langle V_{y}\rangle|$ vs $n$ for a system with $f_{p} = 3.0$. For low
density, $|\langle V_{y}\rangle|$ is small and steadily increasing. 
$|\langle V_{y}\rangle|$ reaches a peak near $n = 0.4$ and
then gradually decreases for higher $n$. This fall off 
in $|\langle V_y\rangle|$ at larger $n$ occurs because
the increasing strength of the particle-particle interactions 
reduces the effectiveness of the pinning, and portions of the
lattice crystallize and become rigid at higher values of $n$. 
In Fig.~9 it was shown that stronger pinning 
increases the magnitude of the maximum value of
$|\langle V_y\rangle|$ for fixed particle density, highlighting the
central role that the pinning plays in the transverse ratchet effect.
Another measure of the reduced effective pinning at high densities 
is shown in Fig.~10(b)
where we plot the dynamic phase diagram as a function of $A$ and $n$
for the same system in Fig.~9. 
The squares mark the separation between the pinned and plastic flow regimes, 
and the circles separate the plastic flow from the ordered regime. 
As $n$ increases, the depinning threshold drops and the pinned phase
monotonically decreases in extent.
The value of $A$ at which the
transition to the ordered phase occurs
is relatively constant as a function of $n$,  
but shows a very small peak feature
similar to what is seen in Fig.~10(a). 
This result indicates that the ratchet effect arises due to 
the collective interactions between the particles, since when 
the interactions dominate over the pinning, the 
transverse ratcheting effect is reduced or destroyed. 

We next consider finite size effects in a system where we set
$n = 0.4$, $f_{p} = 3.0$, and fix the ratio of the number
of particles to the number of pinning sites. 
We perform a series of simulations
for increasing system size $L$ and plot the 
maximum value of  $|\langle V_{y}\rangle|$ vs $L$ in Fig.~10(c). 
We find only a slight decrease in $|\langle V_{y}\rangle|$ at small $L$, 
and observe that $|\langle V_y\rangle|$ is constant at higher $L$.
This indicates that the transverse ratchet effect is not a finite 
size effect and that the value of $|\langle V_{y}\rangle|$ is 
determined by the relative density and strength of the pins.

Next we consider the effect of finite temperature. 
In Fig.~10(d) we plot the maximum value of 
$|\langle V_{y}\rangle|$ vs $T$ for the same system in 
Fig.~9(a) for $n = 0.4$. Here, we find that 
$|\langle V_{y}\rangle|$ fits well to an exponential decay, 
$|\langle V_{y}\rangle| \propto \exp(-\alpha T)$. This
indicates an activated transport mechanism. In the plastic flow regime where 
the transverse ratchet effect occurs, moving particles 
circulate around pinned particles, but as
the temperature increases, the pinned particles become activated 
out of the pinning sites and the ratchet effect is lost. 

We also note that another test of the fact that the pinning density 
and strength controls the effectiveness of the
transverse ratchet is to vary the number of particles while 
holding the number of pinning
sites fixed. As the number of particles increases, the maximum in 
$|\langle V_{y}\rangle|$ stays at roughly 
the same value (not shown) since the number of pinned particles,
which controls the ratchet effect, remains constant.      

\begin{figure}
\center{
\epsfxsize=3.5in
\epsfbox{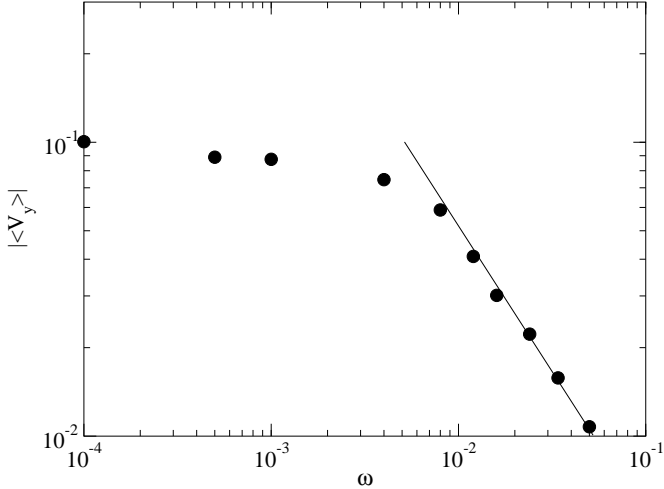}}
\caption{
$|\langle V_{y}\rangle|$ vs $\omega$ for fixed $A = 2.0$, $n = 0.4$ and
$f_{p} = 3.0$. The solid line is a fit to $1/\omega$. 
}
\end{figure}

\begin{figure}
\center{
\epsfxsize=3.5in
\epsfbox{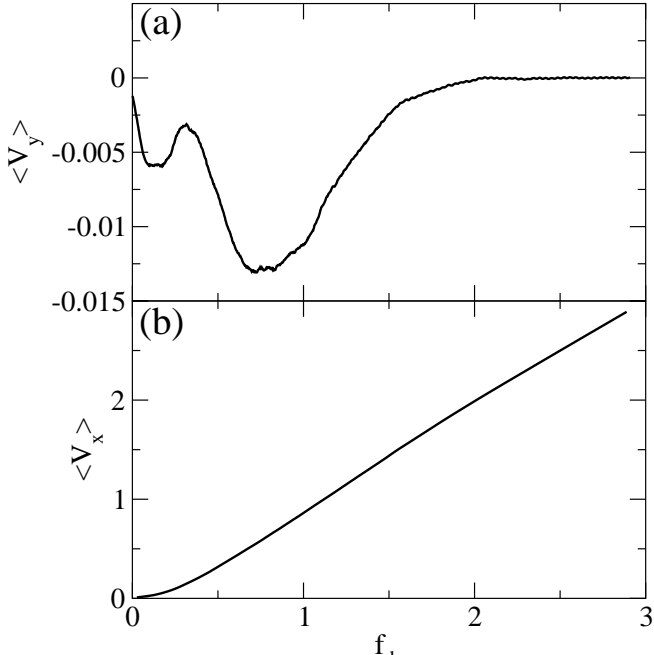}}
\caption{
(a) $\langle V_{y}\rangle$ vs $f_{dc}$ and (b) $\langle V_{x}\rangle$ 
vs $f_{dc}$ 
for $f_{p} = 1.4$, $n = 0.4$, and fixed $A =0.5$. 
}
\end{figure}

\subsection{Effects of Varied Frequency and dc drive}

In Fig.~11 we show the maximum value of $|\langle V_{y}\rangle|$ 
vs frequency $\omega$ for fixed $A = 2.0$ with the same
parameters as in Fig.~10(a) for $n = 0.4$. 
For low frequencies, $|\langle V_{y}\rangle|$ saturates to a constant
value. For the higher frequencies, 
$|\langle V_{y}\rangle|$ decreases approximately as $1/\omega$. We note 
that as the frequency increases, the effective radius of the 
circular particle orbit decreases. 
For increasing $A$ and fixed $\omega$, 

\begin{figure}
\center{
\epsfxsize=3.5in
\epsfbox{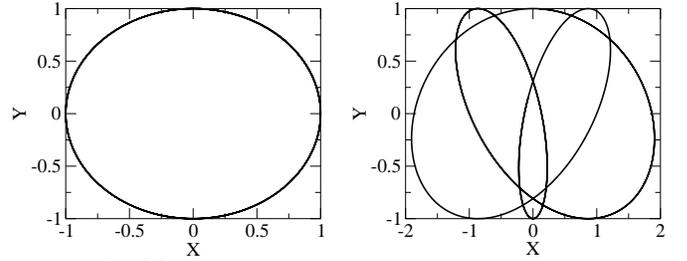}}
\caption{
(a) Orbit for a system with a circular ac drive. (b) Orbit for a 
system with 
$g(t)=\cos(6\omega t){\hat{\bf y}}+[\sin(4\omega t)+\sin(6\omega t)]{\hat {\bf x}}$, with $\omega=0.001$.  
}
\end{figure}

\noindent
the radius of the orbit increases, 
and in Fig.~7 we found that $|\langle V_{y}\rangle|$
increases linearly with $A$.  
From these sets of simulations we find that the 
effective transverse ratchet effect
as measured by the maximum value of 
$|\langle V_{y}\rangle|$ goes as $(A/\omega)f_{p}\exp(-\alpha T)$. Also, as 
a function of $n$, the system shows a maximum 
in $|\langle V_y\rangle|$, indicating that there is an optimal value
of particle-particle interaction strength for the transverse ratchet
effect.

We have also considered the case of fixed $A$ and increasing $f_{dc}$. 
In general we find a similar transverse ratchet effect and dynamical reordered
regime at high $f_{dc}$. 
In this case, $\langle V_{x}\rangle$ 
monotonically increases and there is no overshoot effect.  
In Fig.~12(a) we plot 
$\langle V_{y}\rangle$ vs $f_{dc}$ and in Fig.~12(b) we show
$\langle V_{x}\rangle$ vs $f_{dc}$ for a system with fixed 
$f_{p} = 1.5$, $A = 0.5$, and $n = 0.4$. 
For $f_{dc} > 2.0$, the system reorders to a moving crystal. At low
drives there is some creep due to the application of the ac drive. As
$f_{dc}$ increases, more of the particles become depinned and can ratchet
so a peak in $\langle V_{y}\rangle$ occurs near 
$f_{dc} = 0.9$, while at high drives all particles are depinned
and $\langle V_{y}\rangle$ goes to zero. 
There is an interesting second peak feature 
in $\langle V_y\rangle$ at
low $f_{dc}$. This occurs near what would be the
dc depinning threshold at $A = 0.0$. 
For $f_{p} > 3.0$ this second peak structure disappears.

\section{Ratchet Effects Without dc Drives} 

If we consider a system with $f_{dc} = 0$, 
it is still possible to obtain a nonzero dc response in either the
$x$ or $y$-direction if the closed orbit of the ac drive 
breaks an additional spatial symmetry. 
For particles moving on periodic substrates, 
it was previously shown that a rich variety of 
phase locking and ratchet effects can occur when the ac drive is 
noncircular \cite{Transverse,Guantes}. 
Some of these effects include a number of current
reversals as the ac amplitude is increased \cite{Transverse}. 
In the case of a disordered substrate,
we find that it is still possible to 
produce a ratchet effect and even reversals when $f_{dc}=0$
and an acircular ac drive is applied;
however, the well defined phase locking steps seen for periodic 
substrates are missing. 
In Fig.~13 we show two examples of the closed orbits that a 
particle would follow in the absence
of pinning or other particles. 

\begin{figure}
\center{
\epsfxsize=3.5in
\epsfbox{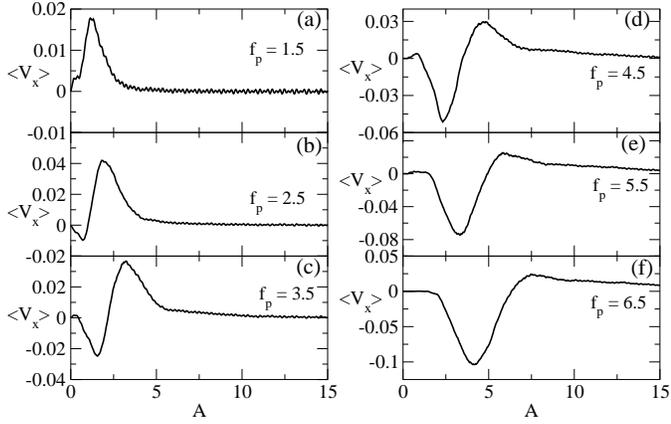}}
\caption{
$\langle V_{x}\rangle$ vs $A$ for a system driven with the ac drive 
shown in Fig.13(b)  
for $f_{p} =$ (a) 1.5, (b) 2.5, (c) 3.5, (d) 4.5, (e) 5.5, and 
(f) $6.5$.
}
\end{figure}

\begin{figure}
\center{
\epsfxsize=3.5in
\epsfbox{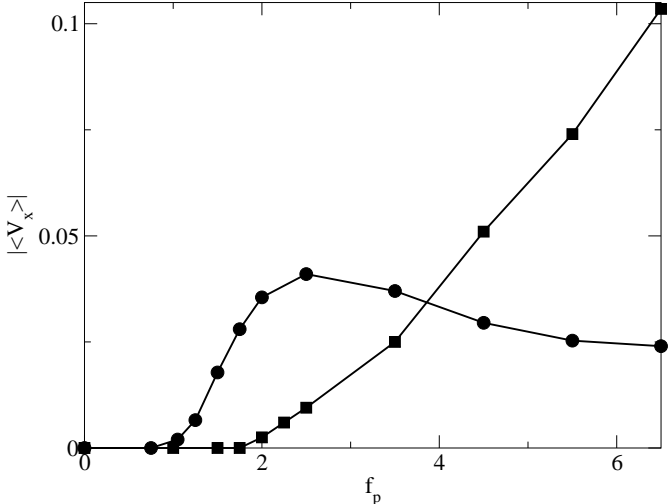}}
\caption{The maximum 
$|\langle V_{x}\rangle|$ for the initial negative peak (squares) and
the positive peak (circles) vs $f_{p}$ for the system shown in Fig.~14. 
}
\end{figure}

\noindent
Figure 13(a) illustrates a circular orbit which does not break
spatial symmetry. 
With only a circular ac drive and no dc drive, the particles do not ratchet.
In Fig.~13(b) we show an orbit for
$g(t)=A\cos(6\omega t){\hat{\bf y}}+A[\sin(4\omega t)+\sin(6\omega t)]{\hat{\bf x}}$,
with $\omega=0.001$. 
Here the reflection symmetry across the $y$ axis is broken. 
In Fig.~14 we plot 
$\langle V_{x}\rangle$ vs $A$ 
for $f_{dc} = 0$, $n = 0.4$, and varied $f_{p}$ for 
the system with the ac drive shown in Fig.~13(b). 
In Fig.~14(a) for low $f_{p} = 1.5$, there is a 
positive peak in $\langle V_{x}\rangle$, indicating
that a $+x$ rectification is occurring even the absence of a dc drive. 
At high $A$, $\langle V_{x}\rangle$ 
goes to zero and the system reorders into a moving crystal. For $f_{p} = 2.5$
[Fig.~14(b)],
the positive peak in $\langle V_x\rangle$ is
larger and there is an additional negative
peak in $\langle V_x\rangle$
that occurs before the positive $\langle V_x\rangle$ peak. As 
$f_{p}$ increases for 

\begin{figure}
\center{
\epsfxsize=3.5in
\epsfbox{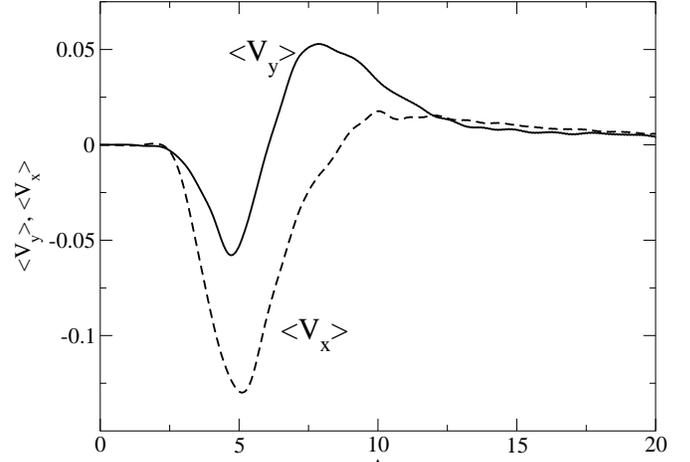}}
\caption{
Ratchet effect in $\langle V_{y}\rangle$ 
(solid curve) and $\langle V_{x}\rangle$ (dashed curve) for the same system in 
Fig.~14 with $f_{p} = 7.0$.  
}
\end{figure}

\noindent
Fig.~14(c-f), the negative peak 
in $\langle V_x\rangle$ 
grows in magnitude and 
the locations of both peaks shift to higher values of $A$. 
For the higher values of $f_{p}$, there is an initial
pinned region where all the particles are trapped and 
$\langle V_{x}\rangle = 0$. 
As $f_{p}$ increases, this pinned region grows in width. 
In Fig.~15 we plot the absolute value of 
the maximum $|\langle V_{x}\rangle|$ for (squares) the initial
negative peak and (circles) the subsequent positive peak. 
Here it can be seen that 
there is a minimum pinning force of $f_{p}\approx 0.9$ 
required to produce any kind of ratcheting effect. 
The positive peak initially grows in magnitude
and reaches a maximum value at $f_p\approx 2.5$.
As $f_p$ increases further, the maximum height of the 
positive peak decreases; however, 
the range of $A$ over which a positive
ratchet effect occurs is also widened as seen in Fig.~14. The negative peak
first appears at a higher value of $f_{p} = 1.75$ and increases monotonically 
in size as $f_{p}$ is further increased.  

We find that the ratchet effect for asymmetric ac drives
can occur in both the $x$ and $y$ 
directions simultaneously.
In Fig.~16 we plot $\langle V_{x}\rangle$ (dashed curve) and 
$\langle V_{y}\rangle$ 
(solid curve) for the same system as in 
Fig.~14 with $f_{p} = 7.0$. Here, the same trend in 
$\langle V_{x}\rangle$ is seen with an initial pinned regime
followed by a strong negative peak and a reversal 
to a positive peak at higher $A$. 
For $\langle V_{y}\rangle$, we find a similar trend,
with an initial negative peak which begins at the 
same value of $A$ as the negative peak in $\langle V_{x}\rangle$. The negative
peak for $\langle V_{y}\rangle$ has a smaller magnitude than 
the negative peak in 
$\langle V_{x}\rangle$. For higher $A$, 
a much larger positive peak in 
$\langle V_{y}\rangle$ occurs near the value of $A$ where 
$\langle V_{x}\rangle$ 
crosses zero. At very high $A$ the ratchet effect in both
directions disappears. 

These results show that it is possible to obtain a ratchet effect 
for particles interacting
with random disorder where the symmetry breaking comes from 
the ac drive alone. The
current reversals and the magnitude of the ratchet 
effect persist for varied system size
and thus they are not artifacts caused by the system size. 
Additionally, if we repeat the simulations
for different random pinning configurations, 
the same types of curves are produced. 
We have also tested a 
number of different ac drives that break a reflection
symmetry. The details
of the curves are different; however, in general at 
least one current reversal is observed in each case.   

The ratchet effect in the absence of the dc drive is 
consistent with the observations
in earlier studies of particles driven over periodic 
disorder \cite{Transverse}. In the previous 
study, crossed ac drives with characteristics similar 
to those shown in Fig.~13(b) produced dc transport
in both the $x$ and $y$ directions. 
In that system the ratcheting effect occurred in well defined
regimes. 
Additionally, the ratchet effect showed a number of current reversals for
increasing ac amplitude; however, for low $A$, 
most of the steps were in the negative direction, while
for higher $A$, most of the steps were in the positive direction. 
This is consistent with the results obtained here for random
disorder where we find a general smearing of all the steps.       
The ratchet effect in both the periodic and random disorder cases does
not occur for ac drives that do not break at least one reflection symmetry.   

\section{Summary}

In summary, we have shown that for repulsively interacting particles 
moving over a disordered substrate, 
it is possible to obtain a transverse ratchet effect when a 
circular ac drive is imposed over an additional dc drive. 
In the absence of the circular ac drive the particles move 
only in the direction of the dc drive. 
When the circular ac drive is present, the particles
can have an additional dc drift velocity in the direction 
transverse to the applied dc drive.
The average transverse velocity shows a peak as a function of 
ac amplitude for fixed dc drive
or for fixed ac amplitude and increasing dc drive. At high ac or dc drives 
the system reorders
to a moving crystal phase. This phase is distinct from a moving smectic, which 
would occur if only a dc drive were applied. 
In the reordered phase the transverse ratchet effect is lost. For fixed 
dc drive and increasing ac drive we also observe what we term 
an overshoot effect where the longitudinal 
velocity is larger than the maximum possible value that could be produced
by the dc drive alone.  This
overshoot effect arises due to a longitudinal ratchet effect 
produced near the reordering transition where the 
nonlinearity of the longitudinal velocity force response changes. We 
analyze the transverse ratchet effect for a wide range of system parameters, 
including disorder strength,
particle density, temperature, and ac frequency. 
We also show that the ratchet effect is robust for increasing
system sizes. The transverse ratchet effect arises due to a symmetry 
breaking by the chirality of the ac drive plus an additional symmetry 
breaking by the dc drive. 
If the crossed ac drives are more complicated, such that the closed orbits 
themselves are asymmetric, then it is possible to obtain a ratchet 
effect even in the absence of a dc drive.       

Acknowledgments---We thank Z. Toroczkai and M.B. Hastings 
for useful discussions. 
This work was supported by the US DOE under Contract No. W-7405-ENG-36.

\end{document}